\documentstyle[11pt]{article}
\date{}

\begin{document}
\title{A New Mass Formula for NG Bosons in QCD}
\author{{Reginald T. Cahill  and      Susan M. Gunner
  \thanks{E-mail: Reg.Cahill@flinders.edu.au,
ECSMG@psy1.ssn.flinders.edu.au}}\\
  {Department of Physics, Flinders University}\\ { GPO Box 2100, Adelaide 5001,
Australia}\\
{July 1995 - hep-ph/9507268} }
\maketitle
\begin{center}
\begin{minipage}{120mm}
\vskip 0.6in
\begin{center}{\bf Abstract}\end{center}
{An often used mass formula for Nambu-Goldstone (NG)  bosons in QCD, such as
the pions, involves
the condensate $<\overline{q}q>$, $f_{\pi}$ and the quark current masses.  We
argue, within the
context of the Global Colour Model to QCD, that this expression is wrong.
Analysis of the
interplay between  the Dyson-Schwinger equation for the constituent quark
effect and the
Bethe-Salpeter equation for the NG boson results in a new mass formula.

PACS numbers: 12.38.Lg, 11.30.Qc, 11.30.Rd, 14.40.Aq, 11.10.St, 12.38.Aw

Keywords:  PCAC, Nambu-Goldstone boson, pion  mass formula, Quantum
Chromodynamics}
\end{minipage} \end{center}

\vskip 1.5cm
\par

The computation of the low energy properties of QCD is a difficult
non-perturbative problem
in quantum field theory.  However one species of hadron, the (almost)
Nambu-Goldstone (NG)
bosons, such as the pions, have always played a key role. Because they are
directly associated
with the dynamical breaking of chiral symmetry their properties are strong
indicators of the
nature of the underlying quark-gluon dynamics in QCD.  However  the   small
current masses of the
u and d quarks means that the pions are not strictly massless, as for  true NG
bosons, but
acquire small masses. They are therefore also significant as low mass hadronic
excitations.

One expects that there should be some perturbative expression for   the pion
mass in terms
of the quark current masses  which   is built upon the underlying
non-perturbative
chiral-limit quark-gluon dynamics. While the relation of the low pion mass to
the breaking of
chiral symmetry dates back to the current algebra era and PCAC \cite{GMOR}, the
often used
implementation in QCD has the form,
\begin{equation}
M_{\pi}^2= \frac{(m_u+m_d)\rho}{f_{\pi}^2}
\end{equation}
where the integral $\rho=<\overline{q}q>$ is the so called condensate
parameter. For $N_c=3$
\begin{equation}
\rho=N_ctr(G(x=0))=12\int\frac{d^4q}{(2\pi)^4}\sigma_s(q^2),
\end{equation}
and $f_{\pi}$ is the usual pion decay constant.
In (2)  $\sigma_s(s)$   is the chiral limit ($ m \rightarrow 0 $) scalar part
of the
constituent quark propagator   \begin{equation}
G(q)=(iA(s;m)q.\gamma+B(s;m)+m)^{-1}=-iq.\gamma\sigma_v(s;m)+\sigma_s(s;m).
\end{equation}

We note  that the expression for $\rho$ in (2) is
divergent  in QCD, because for large $s \rightarrow \infty$
$B(s)$ decreases like  $1/s\mbox{ln}[s/\Lambda^2]^{1-\lambda}$
where $\lambda = 12/(33-2N_f)$ and $\Lambda$ is the QCD scale parameter.   Some
integration cutoff  is usually
 introduced.   The  values of $m$ and
$<\overline{q}q>$ are then usually quoted as being relative to  some cutoff
momentum, often $1
GeV$. An alternative approach  \cite{Bijnens} is to use finite energy sum rules
and Laplace sum
rules.

Here we present a new analysis of the chiral symmetry breaking in the Global
Colour Model (GCM)
to QCD. We extract a new expression for the pion mass, which essentially
replaces (2).
Nevertheless the new result, see equation (22) below, is very similar to (1)
except that it
contains a naturally arising cutoff function $c(s)$, and also a dynamical
enhancement function
$\epsilon_s(s)$ for  the quark current mass $m$. There is also a factor of 2
difference. In the
Appendix we give an example of the type of incorrect analysis that leads to the
form in (2). The
new mass formula  implies that the pion mass is dominated by IR processes, and
not UV processes
as in (2).

An overview and an insight into the nature of the non-perturbative low energy
hadronic regime
of  QCD is provided by the functional integral hadronization of QCD
\cite{RTC,Reinhardt90}.
This amounts to a dynamically determined change of functional integration
variables, from
quarks and gluons, to bare hadrons

$
\int {\cal D}\overline{q}{\cal D}q{\cal
D}Aexp(-S_{qcd}[A,\overline{q},q]+\overline{\eta}q+ \overline{q}\eta)\approx$
\begin{equation} \int{\cal
D}\pi{\cal D}\overline{N}{\cal
D}N...exp(-S_{had}[\pi,...,\overline{N},N,..]+J_{\pi}[\overline{\eta},\eta]\pi+..)
\end{equation}

  The final functional integration over the hadrons  gives the
hadronic observables, and amounts to dressing each hadron by, mainly, lighter
mesons.  This
functional integral transformation cannot  yet be done exactly.  The basic
insight is that the
quark-gluon dynamics, on the LHS of (4),  is fluctuation dominated, whereas the
RHS is not, and
for example the meson dressing of bare hadrons is known to be almost
perturbative. In performing
the change of variables essentially normal mode  techniques are used
\cite{RTC}. In practice this
requires detailed numerical computation of the gluon propagator, quark
propagators, and  meson
and baryon propagators. The mass-shell states of the latter are determined by
covariant
Bethe-Salpeter  and  Faddeev equations. The Faddeev computations are made
feasible by using the
diquark correlation propagators, which must also be determined.

 The first and easiest formal
transformation results from doing the gluon integrations, leaving an action for
quarks of the
form
 \begin{equation}S[\overline{q},q]= \int
\overline{q}(x)(-\gamma . \partial+{\cal M}) q(x) +\frac{1}{2}\int
j^a_{\mu}(x)j^a_{\nu}(y)D_{\mu\nu}(x-y)+\frac{1}{3!}\int
j^a_{\mu}j^b_{\nu}j^c_{\rho}D^{abc}_{\mu\nu\rho}+...
\end{equation}
where $j^a_{\mu}(x)=\overline{q}(x)\frac{\lambda^a}{2}\gamma_{\mu}q(x)$.
The GCM  is a model field theory for QCD  based on a  truncation
of $S[\overline{q},q]$ in which
 the  higher order n-point ($n\geq 3$) functions are neglected, and   only the
gluon 2-point
function  $D_{\mu\nu}(x-y)$  is retained.

The GCM is thus a quantum field theory that can also  be considered to be
defined by the action
$$
S_{gcm}[\overline{q},q,A^a_{\mu}]=\int \left(
\overline{q}(x)(-\gamma . \partial+{\cal
M}+iA^a_{\mu}\frac{\lambda^a}{2}\gamma_{\mu})\delta(x-y)q(y) +\right.$$
\begin{equation}
\left. +\frac{1}{2}
A^a_{\mu}(x)D^{-1}_{\mu\nu}(i\partial)\delta(x-y)A^a_{\nu}(y) \right)
\end{equation}
where the matrix $D^{-1}_{\mu\nu}(p)$  is the inverse of $D_{\mu\nu}(p)$, which
in turn is the
Fourier transform of $D_{\mu\nu}(x)$. This action has a
global colour symmetry. The GCM is thus analogous to QED except for  colour
currents and
the non-quadratic phenomenological form for $D^{-1}_{\mu\nu}(p)$  in the pure
gluon sector.
The determination of the best-fit phenomenological form for  $D_{\mu\nu}(p)$
was reported in
\cite{CG1} using  the  separable expansion technique to facilitate numerical
computations and
the fitting to some meson data.

Having made the GCM truncation in (5) it is possible to proceed further and
to transform \cite{RTC} the quark functional
integrations into the hadronic functional integrations, as in (4). If the
additional approximation
$D_{\mu\nu}(x-y)\rightarrow g\delta_{\mu\nu}\delta(x-y)$ is made in (6), i.e. a
contact coupling of the quark currents, then the NJL type models are obtained.
If in (4)
a derivative expansion of the complete non-local  hadronic effective action is
performed, then the
Chiral Perturbation Theory (CPT) phenomenology is obtained. However in the GCM,
with appropriate
$D_{\mu\nu}(x)$, all computations are finite and no cutoffs or renormalisation
procedures are used. As
well, using a mean field approximation, the soliton phenomenology for the
baryons may be derived
\cite{CR85}, and has been studied in
\cite{Frank}.

In analogy with QED the GCM involves the determination of various coupled
equations for the
propagators. In the approximation often employed (which is motivated by the
functional
integration  analysis of (6), see \cite{RTC}) the first equation is  the
Dyson-Schwinger (DSE)
equation for the constituent quark propagator (the so-called vacuum equation of
the GCM
\cite{CR85,RTC}),

\begin{equation}
B(p^2;m)=\frac{16}{3}\int\frac{d^4q}{(2\pi)^4}D(p-q).\frac{B(q^2;m)+m}{q^2A(q^2;m)^2+(B(q^2;m)+m)^2},
\end{equation}

\begin{equation}
[A(p^2;m)-1]p^2=\frac{8}{3}\int\frac{d^4q}{(2\pi)^4}q.pD(p-q).\frac{A(q^2;m)}{q^2A(q^2;m)^2+(B(q^2;m)+m)^2},
\end{equation}
using a Feynman-like gauge, $D_{\mu\nu}(p)=\delta_{\mu\nu}D(p)$, and  the
perturbative quark-gluon
vertex function. The Landau gauge can also be used; see \cite{CG1} for
comparison.

Using Fourier transforms (7) may be written in the form, here for $m=0$,
\begin{equation}
D(x)=\frac{3}{16}\frac{B(x)}{\sigma_s(x)},
\end{equation}
which implies that knowledge of the quark propagator determines the effective
GCM gluon
propagator. Multiplying (9) by $B(x)/D(x)$, and using  Parseval's identity for
the RHS, we
obtain the identity
\begin{equation}
\int d^4x
\frac{B(x)^2}{D(x)}=\frac{16}{3}\int\frac{d^4q}{(2\pi)^4}B(q)\sigma_s(q).
\end{equation}

The second basic equation is the Bethe-Salpeter equation (BSE) for the pion
mass-shell state at
the level of approximation that matches (7) and (8) in the GCM analysis
\cite{RTC,CRP87}
\begin{equation}
\Gamma^f(p,P)=\frac{8}{3}\int\frac{d^4q}{(2\pi)^4}D(p-q)tr_{SF}(G_+T^gG_-T^f)\Gamma^g(q,P)
\end{equation}
where $G_{\pm}=G(\pm q-\frac{P}{2})$. This BSE is for isovector NG bosons, and
only the dominant
$\Gamma=\Gamma^f T^f i\gamma_5$ amplitude is retained (see \cite{CRP87} for
discussion); the spin
trace arises from projecting onto this dominant amplitude. Here
$\{T^b,b=1,..,N_F^2-1\}$ are the
generators of $SU(N_F)$, with $tr(T^fT^g)=\frac{1}{2}\delta_{fg}$.

The BSE (11) is an implicit
equation for the mass shell $P^2=-M^2$.  It  has solutions {\em only} in
the time-like region $P^2 \leq 0$. Fundamentally this is ensured by (7) and (8)
being the
specification of an absolute minima of an effective action after a bosonisation
\cite {RTC}.
Nevertheless the loop momentum  is kept in the
space-like region $q^2 \geq 0$; this mixed metric device ensures that the
quark and gluon propagators remain close to the real space-like region where
they have been most
thoroughly studied. Very little is known about these propagators in the
time-like region $q^2 <
0$. The GCM gives a detailed
description of the pion properties, including its coupling to other states, in
the language of
effective non-local actions \cite{RTC}.

The non-perturbative quark-gluon dynamics is expressed here in (7) and (8).
Even when  $m=0$
eqn.(7) can have non-perturbative solutions with $B\neq 0$. This is the
dynamical breaking of
chiral symmetry.

 When $m=0$ eqn.(11) has a solution for $P^2=0$; the Goldstone theorem effect.
For the zero linear momentum  state  $\{P_0=0,\vec{P}=\vec{0}\}$  it is easily
seen that eqn.(11)
reduces to eqn.(7) with $\Gamma^f(q,0)=B(q^2)$.  When $\vec{P}\neq \vec{0}
\mbox{\ \ } \mbox{\
then \ }  \Gamma^f(q,P)\neq B(q)$, and (11)  must be solved  for
$\Gamma^f(q,P)$.

We shall now determine an accurate expression for the mass of the pion when $m
\neq 0$. This
amounts to finding an analytic solution to the BSE (11), when the constituent
quark propagators
are determined by (7) and (8). The result will be accurate to order $m$.

 For small $m \neq 0$
we can introduce the Taylor expansions in $m$
\begin{equation}
B(s;m)+m = B(s)+ m.\epsilon_s(s)+O(m^2),
\end{equation}
\begin{equation}
A(s,m) = A(s)+m.\epsilon_v(s)+O(m^2).
\end{equation}
 For large space-like $s$ we find that $\epsilon_s
\rightarrow 1$, but for small $s$ we find that $\epsilon_s(s)$ can be
significantly larger than 1
(see Fig.1). This is a dynamical enhancement  of the quark current mass by
gluon dressing in the
infrared region. Even in the chiral limit  the quark running mass
$M(s)=B(s)/A(s)$  is
essential for understanding any non-perturbative QCD quark effects. At $s=0.3
GeV^2$ we find
\cite{CG1} that $M(s) \approx 270MeV$.

Because the pion mass $M_{\pi}$ is small when $m$ is small, we can perform an
expansion of the
$P_{\mu}$ dependence in the  kernel of (11). Since the analysis is Lorentz
covariant we can,
without loss of validity, choose to work in the rest frame with
$P=(iM_{\pi},\vec{0})$,
giving, for  equal mass quarks for simplicity, $$
\Gamma(p)=\frac{2}{9}M_{\pi}^2\int\frac{d^4q}{(2\pi)^4}D(p-q)I(s)\Gamma(q)+
\hspace{80mm} $$
\begin{equation}
\mbox{\ \ \ \ \ \ \ \ \ }
+\frac{16}{3}\int\frac{d^4q}{(2\pi)^4}D(p-q)\frac{1}{s(A(s)+\epsilon_v(s).m)^2+(B(s)+
m.\epsilon_s(s))^2}\Gamma(q)+....,
\end{equation}
where
\begin{equation}
I(s)=6\left(\sigma_v^2-2(\sigma_s\sigma_s'+s\sigma_v\sigma_v')-s(\sigma_s\sigma_s''-(\sigma_s')^2)
-s^2(\sigma_v\sigma_v''-(\sigma_v')^2\right).
\end{equation}

By using Fourier transforms the  integral equation (14), now with explicit
dependence on
$M_{\pi}$,  can be expressed in the form of a variational mass functional,
\begin{equation}
M_{\pi}[\Gamma]^2=-\frac{24}{f_{\pi}[\Gamma]^2}\int\frac{d^4q}{(2\pi)^4}\frac{\Gamma(q)^2}
{s(A(s)+\epsilon_v(s).m)^2+(B(s)+
m.\epsilon(s))^2}+\frac{9}{2f_{\pi}[\Gamma]^2}\int d^4x\frac{\Gamma(x)^2}{D(x)}
\end{equation}
in which
\begin{equation}
f_{\pi}[\Gamma]^2 =
\int\frac{d^4q}{(2\pi)^4}I(s)\Gamma(q)^2.
\end{equation}

The functional derivative $\delta M_{\pi}[\Gamma]^2/\delta\Gamma(q)=0$
reproduces (14).
The mass functional  (16) and its  minimisation is equivalent to the pion BSE
in the near chiral
limit.  To find an estimate for the minimum we need only note that the change
in
$M_{\pi}^2$ from its chiral limit value of zero will be of 1st order in $m$,
while the change in
the zero linear momentum frame $\Gamma(q)$ from its chiral limit  value
$B(q^2)$ will be of 2nd
order in $m$.

Hence to lowest order in $m$ we have that the pion mass is given by
$$
M_{\pi}^2=\frac{48m}{f_{\pi}[B]^2}\int\frac{d^4q}{(2\pi)^4}\frac{\epsilon_s(s)B(s)+
s\epsilon_v(s)A(s)}{sA(s)^2+B(s)^2}
\frac{B(s)^2}{sA(s)^2+B(s)^2}
\mbox{\ \ \ \ \ \ \ \ \ \ \ \ \ \ \ \ \ \ \ \ \ \ \ \ \ \ \ \ \ \ \ }$$
\begin{equation}
-\frac{24}{f_{\pi}[B]^2}\int\frac{d^4q}{(2\pi)^4}
\frac{B(s)^2}{sA(s)^2+B(s)^2}+\frac{9}{2f_{\pi}[B]^2}\int
d^4x\frac{B(x)^2}{D(x)}+O(m^2)
\end{equation}
However the pion mass has been shown to be zero  in the chiral limit. This is
confirmed
as the  two $O(m^0)$ terms in (18)  cancel because of the identity (10).  Note
that it might
appear that $f_{\pi}$ would contribute an extra $m$ dependence from its kernel
in (15).  However
because the numerator in (16) is already of order $m$, this extra contribution
must be of higher
order in $m$.

Hence we  finally arrive at the  analytic expression, to $O(m)$, for the NG
boson
$(\mbox{mass})^2$ from the solution of the BSE in (11) which includes the
non-perturbative gluon
dressing to give  constituent quarks
\begin{equation}
M_{\pi}^2=\frac{2m\rho_{eff}}{f_{\pi}^2}
\end{equation}
where
\begin{equation}
\rho_{eff}=24\int\frac{d^4q}{(2\pi)^4}\left(\epsilon_s(s)
\sigma_s(s)+s\epsilon_v(s)
\sigma_v(s)\right)c(s)
\end{equation}
defines an effective condensate parameter  $\rho_{eff}$. The  $\rho_{eff}$
integrand involves a
naturally arising function \begin{equation}
c(s)=\frac{B(s)^2}{sA(s)^2+B(s)^2},
\end{equation}
which acts as a smooth cutoff function. It is this function which causes the
pion mass to be
IR dominated. Of course the determination of $B(s)$ and $A(s)$  by (7) and (8)
involves both IR
and UV momenta. As well $\rho_{eff}$ contains a contribution from the vector
part of the
chiral-limit quark propagator. Because this vector part makes a small
contribution we can
approximate (20) by the form
\begin{equation}
\rho_{eff}=24\int\frac{d^4q}{(2\pi)^4}\epsilon_s(s)c(s)\sigma_s(s)
\end{equation}
This result contradicts (2).
Significantly the  factor $c(s)$  makes (22) convergent when $B(s)$ has the QCD
determined
asymptotic form discussed  after (3). The $\epsilon$ enhancement function is
important to the
numerical values but does not affect the convergence analysis because
asymptotically
$lim_{s \rightarrow \infty}\epsilon_s(s) = 1$.

We now give examples of the functions occuring in (22). In ref. \cite{CG1} the
GCM
phenomenological chiral-limit  quark propagator and the corresponding gluon
propagator were
obtained by fitting a number of  meson observables to meson data.
{}From these forms the cutoff function $c(s)$ and  the enhancement
function $\epsilon_s(s)$ can be computed.  The resulting
functions are  shown in Fig.1.  We note that the value of  $\epsilon_s(s)
\approx 4$ implies that in the infrared region, appropriate to the internal
dynamics of hadrons,
the quark current mass of $\sim 6 MeV$ is enhanced by gluon dressing to some
$24MeV$. Of course
the major effect is the chiral-limit constituent mass of some $270MeV$.
 We compute using (22) that $\rho_{eff}=(0.233 GeV)^3$, and from (17) that
$f_{\pi}=93.0MeV$. With
$m=6.5MeV$ (22) gives  $M_{\pi}=138.5MeV$ without cutoffs or renormalisation
procedures.
We note that these numerical results are actually close to the usual values
using the forms
in (1) and (2) with the traditional cutoff of $1 GeV$. This may be incidental
to the particular
parameterisation  used for the phenomenologcal GCM propagators used in
\cite{CG1}.

 It would be
interesting to see if the corrected form in (22) can be  rigorously obtained
within QCD.

\noindent {\bf Appendix:} Here we present one form of the flawed analysis that
leads to the
incorrect result in (2). Consider the mass functional in (16). Suppose we make
the ansatz that
\begin{equation}
\Gamma_{\pi}(q) = B(q^2;m).
\end{equation}
{}From (7) we obtain by Fourier transforms and Parseval's identity, in analogy
to (10),
\begin{equation}
\int d^4x \frac{B(x;m)^2}{D(x)}=\frac{16}{3}\int\frac{d^4q}{(2\pi)^4}B(q;m)
\frac{B(q;m)+m}{q^2A(q;m)^2+(B(q;m)+m)^2}.
\end{equation}
Using (23) and (24)  the mass functional (16) then gives
$$
M_{\pi}^2=-\frac{24}{f_{\pi}^2}\int\frac{d^4q}{(2\pi)^4}\frac{B(q;m)^2}
{s(A(s)+\epsilon_v(s).m)^2+(B(s)+
m.\epsilon(s))^2}+
$$
\begin{equation}\frac{24}{f_{\pi}^2}\int\frac{d^4q}{(2\pi)^4}
\frac{B(q;m)(B(q;m)+m)}{q^2A(q;m)^2+(B(q;m)+m)^2}.
\end{equation}
\begin{equation}
=\frac{24}{f_{\pi}^2}\int\frac{d^4q}{(2\pi)^4}\frac{B(q;m)m}
{q^2A(q;m)^2+(B(q;m)+m)^2}.
\end{equation}
\begin{equation}
=\frac{24}{f_{\pi}^2}\int\frac{d^4q}{(2\pi)^4}\frac{B(q)m}
{q^2A(q)^2+(B(q))^2}.
\end{equation}
to $O(m)$, and  this gives eqns(1) and (2). The flaw in this argument is to
overlook
that  in perturbation theory  eigenvalues  change at 1st order and eigenvectors
change at 2nd
order of the perturbation parameter. The ansatz (23) actually gives
$\Gamma_{\pi}(q)$ a hidden
1st order dependence on $m$, and is thus incorrect. Essentially the same error
has been made in
the  functional integral formulation of the GCM, where the ansatz (23) arises
through the choice
of the non-local meson fluctuation. In all GCM papers $\rho$ should be replaced
by $\rho_{eff}$.

RTC acknowledges useful conversations with C. J. Burden, C. D. Roberts, P. C.
Tandy and M.
Thomson. C. D. Roberts \cite{Roberts95} has independently arrived at similar
results
to those reported here.

\vspace{10mm}

{\bf Figure  Caption}

Fig.1 Shows (a) the scalar enhancement function $\epsilon_s(s)$, (b) the
integrand
$s\epsilon_s(s)c(s)\sigma_s(s)$ (from eqn.(22)) of the effective condensate
parameter
$\rho_{eff}$ (in arbitrary units), and (c) the cutoff function $c(s)$.

\begin{figure} [h]
\vspace{110mm}
\hspace{25mm}\special{illustration MathFig.ps}
\end{figure}


\newpage
\begin{thebibliography}{99}

\bibitem{GMOR} M. Gell-Mann, R. Oakes, and B. Renner, Phys. Rev. {\bf
175}(1968)2195.

\bibitem{Bijnens} J. Bijnens, J. Prades and E. de Rafael, Phys. Lett. B {\bf
348}(1995)226.

\bibitem{CR85} R. T. Cahill and C. D. Roberts, Phys. Rev. D {\bf 32}(1985)2419.

\bibitem{RTC} R. T. Cahill,  Nucl. Phys. A {\bf 543}(1992)63c.

\bibitem{Reinhardt90} H. Reinhardt, Phys. Lett. B {\bf 244}(1990)316.

\bibitem{Frank} M. R. Frank and P. C. Tandy, Phys. Rev. C {\bf 46}(1992)338.

\bibitem{CG1}  R. T. Cahill and S. M. Gunner, {\em Quark and gluon propagators
from meson data},
            hep-ph/9506445.

\bibitem{CRP87} R. T. Cahill, C. D. Roberts and J. Praschifka, Phys. Rev. D
{\bf 36}(1987)2804.


\bibitem{Roberts95}  C. D. Roberts, private communication.

\end{thebibliography}
\end{document}